\documentclass[10pt,draft]{dis03}
\usepackage{epsf,amsmath}
\usepackage{amssymb,epsfig,floatflt}

\newcommand{\rp}{$\not$\kern-0.pt$R_p$}

\def\z0{Z^0}

\def\kg{\kappa_{tu\gamma}}

\def\vz{v_{tuZ}}

\begin{document}
\title{SEARCH FOR SINGLE-TOP PRODUCTION\\
WITH THE ZEUS DETECTOR AT HERA\thanks{Talk presented at DIS 2003, 
April 23-27, St. Petersburg, 
Russia}}

\author{Dominik Dannheim (for the ZEUS collaboration)\\
DESY~/~University of Hamburg \\
Notkestr. 85, D-22607 Hamburg, Germany\\
E-mail: dominik.dannheim@desy.de \\}

\maketitle

\begin{abstract}
\noindent A search for anomalous single-top production, $ep\rightarrow etX$, 
$t\rightarrow Wb$, has been made with the ZEUS detector at HERA 
using an integrated 
luminosity of 130~pb$^{-1}$. No evidence for top-quark production
was found in either the leptonic or the hadronic decay channel
of the $W$ boson. The resulting constraints on single-top
production via quark-flavour changing neutral current (FCNC) 
transitions exclude a substantial region in the FCNC $tu\gamma$
coupling not ruled out by other experiments.
\end{abstract}

\section{Single-top production at HERA} 

In the Standard Model (SM), flavour changing neutral current (FCNC) interactions
are suppressed by the GIM mechanism~\cite{GIM}. 
Due to the large mass
of the top-quark, close to the electroweak symmetry breaking scale, deviations
from the SM are expected to be observed first in the top sector.

At the HERA $ep$ collider, with its maximum centre-of-mass energy of $\sqrt{s}=318$~GeV, 
top quarks can only be singly produced. At tree level, the production proceeds through
the charged current (CC) reaction $ep\rightarrow \nu t \bar{b}X$.
Since the SM cross section for single-top production is less than 1~fb, 
any observed single-top event in the present data would be a clear sign of physics 
beyond the SM. The FCNC coupling, $tuV$ or $tcV$, would induce the neutral current (NC)
reaction $ep\rightarrow etX$ \cite{frihol,belyaev}, in which the incoming lepton exchanges a
$\gamma$ or $Z^0$ with an up-type quark in the proton, yielding a top quark in the final 
state (see Fig.~\ref{fig-stop}). 
The $Z^0$ exchange is suppressed by the large propagator mass. Furthermore,
large values of $x\gtrsim0.3$, where $x$ denotes the fraction of the proton momentum
carried by the struck quark, are needed to produce a top quark. Since the $u$-quark
density of the proton is much higher than the $c$-quark density, the production
of single top quarks is most sensitive to a coupling of the type $tu\gamma$.

The search for single-top quark production~\cite{zeussingtop} reported in the 
following sections was based on the complete HERA~I dataset of 130~pb$^{-1}$.

\section{Single-top signatures}
The SM decay, $t\rightarrow bW$, with subsequent leptonic decay of the 
$W$ boson, $W\rightarrow e\nu _e,\mu \nu _\mu$
(11\% branching ratio per channel), leads to the presence
of an isolated high-energy lepton, significant missing transverse momentum
arising from the undetected neutrino and a large value of the hadronic transverse momentum,
$p_T^\mathrm{had}$, stemming from the $b$-quark 
decay. Production of single $W$ bosons
with subsequent leptonic decay is the only SM 
process with a measurable cross section (about 1 pb, evaluated including 
QCD corrections $\mathcal{O}(\alpha^2 \alpha_s)$~\cite{wnlo}),
which leads, at HERA, to events with an isolated lepton and missing 
transverse momentum in the final state. However a steeply falling $p_T^\mathrm{had}$-spectrum
is expected for events from single $W$ production.

In the hadronic decay channel of the $W$ boson, $W\rightarrow q \bar{q'}$ 
(68 \% branching ratio),
three jets are expected in the final state, with the dijet invariant-mass
distribution for the correct pair of jets peaking at the mass of the $W$ boson,
$M_W$, and the three-jet invariant-mass distribution peaking at the mass of 
the top quark, $M_\mathrm{top}$.
QCD multijet events are the main SM background in the hadronic channel.

Anomalous FCNC couplings could lead to the decays
$t\rightarrow u\gamma$ and $t\rightarrow uZ^0$, resulting in multi-jet signatures
with lepton pairs from the decay of the gauge boson.

\section{Leptonic channel}
The main selection requirements for the search in the leptonic decay 
channels of the $W$ boson were large missing transverse momentum as measured
with the calorimeter, $p_T^\mathrm{CAL}>20$~GeV, and the presence of 
an isolated track, 
which had to be 
identified as either electron or muon.
The events had to contain at least one jet.
Events with an electron candidate in a back-to-back configuration
with the hadronic system were rejected, thus reducing background from 
neutral current deep inelastic scattering (NC DIS) events.

The selected data sample contained 24 (12) electron (muon) 
candidate events, in good agreement with $20.6^{+1.7}_{-4.6}$ ($11.9^{+0.6}_{-0.7}$) 
events expected from SM processes.
The SM expectation
in the electron channel is dominated by badly reconstructed 
NC DIS events, while in the muon channel the main contribution is from
Bethe-Heitler 
muon pair production, where only one of the two muons
is detected and the missing transverse momentum is due to mismeasurement.

A subsequent final selection for single-top candidates was applied to the 
preselection of isolated lepton events. The cuts were
optimised using simulations of both the SM background and the expected
single-top signal. In the electron channel, NC DIS background was
further reduced based on the energy-momentum balance. In the muon channel,
a cut on the missing transverse momentum, after correcting for the muon-track 
momentum, was applied to reduce background from Bethe-Heitler muon pair
production.
In both channels, a large value of the hadronic transverse momentum, $p_T^\mathrm{had}>40$~GeV,
was required.
After applying these additional cuts, no events remained in the data sample,
while $0.94^{+0.11}_{-0.10}$ ($0.95^{+0.14}_{-0.10}$) electron (muon)
events were expected from SM processes, mainly from single
$W$ production. 
The efficiency for single-top production was 34\% (33\%) in the electron
(muon) channel.
Table~\ref{tab-yields} summarizes the observed and expected events 
for different stages of the single-top search.
\begin{table}[hbt!]
\small
\centering
\begin{tabular}{|l||c|c|} \hline
 & Electron & Muon \\
 \multicolumn{1}{|c||}{Leptonic channel}  & channel & channel \\ 
 & obs./expected ($W$) & obs./expected ($W$) \\ \hline \hline
Preselection & 24 / $20.6_{-4.6}^{+1.7}\ (17\%)$ & 12 / $11.9_{-0.7}^{+0.6}\ (16\%)$ \\ \hline
Final sel. ($p_T^{\rm had}>25$ GeV) & 
2 / $2.90_{-0.32}^{+0.59}\ (45\%)$ & 5 / $2.75_{-0.21}^{+0.21}\ (50\%)$ \\ \hline
Final sel. ($p_T^{\rm had}>40$ GeV) & 
0 / $0.94_{-0.10}^{+0.11}\ (61\%)$ & 0 / $0.95_{-0.10}^{+0.14}\ (61\%)$ \\ \hline
\end{tabular}
\caption{Number of observed events at different selection stages of the single-top search, 
compared to the SM expectations.
The percentage of single-$W$ production included
in the expectation is indicated in parentheses. The statistical and systematic 
uncertainties added in quadrature are also indicated.}
\label{tab-yields}
\end{table}

\section{Hadronic channel}
The event selection in the hadronic decay channel of the $W$ boson required 
three jets with $p_T^\mathrm{jet1}>40$ GeV,
$p_T^\mathrm{jet2}>25$ GeV and $p_T^\mathrm{jet3}>14$ GeV. 
The photoproduction background was reduced by requiring one of the 2-jet 
masses and the 3-jet mass to be compatible with the $W$ boson mass and the 
top mass, respectively. Events from NC DIS processes were suppressed by
rejecting events with identified electron candidates. 

The normalisation uncertainties of the photoproduction
background, which was only evaluated in leading order of QCD, were reduced by 
normalising the simulated event rates 
to those observed in the low-mass domain. There were 14 events observed in the data,
while $17.6^{+1.8}_{-1.2}$  events were expected from SM background. 
The signal efficiency was 24\% for the hadronic decays of the $W$ boson.

\section{Exclusion limits on FCNC couplings}
\begin{figure}[hbt!]
\psfull
\begin{center}
\begin{minipage}[t]{0.35\linewidth}
\epsfig{figure = 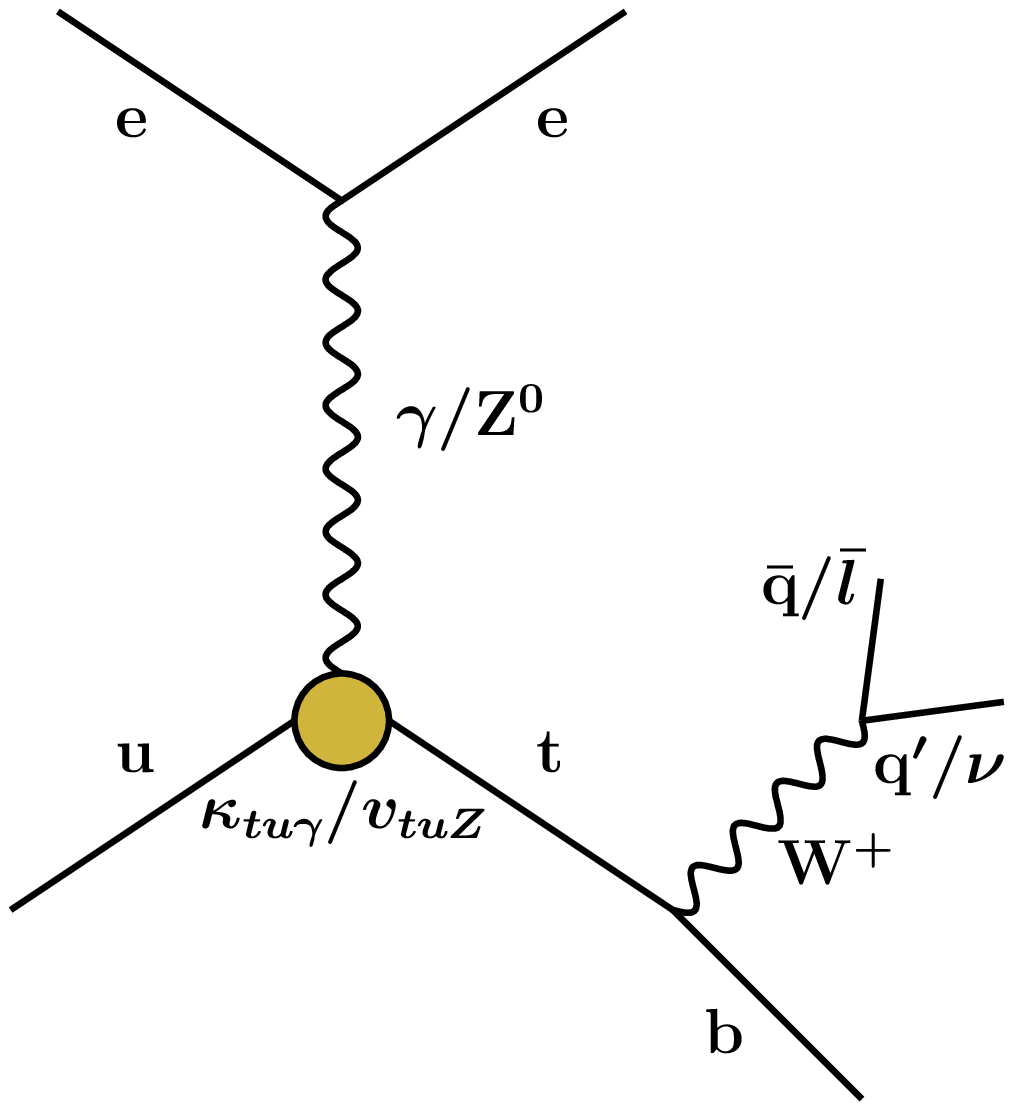,
width={4.2cm}, angle=0, clip=}
\caption[*]{Single top-quark production via FCNC transitions at HERA.}
\label{fig-stop}
\end{minipage}\hfill
\begin{minipage}[t]{0.03\linewidth}
\end{minipage}
\begin{minipage}[t]{0.62\linewidth}
\epsfig{figure =  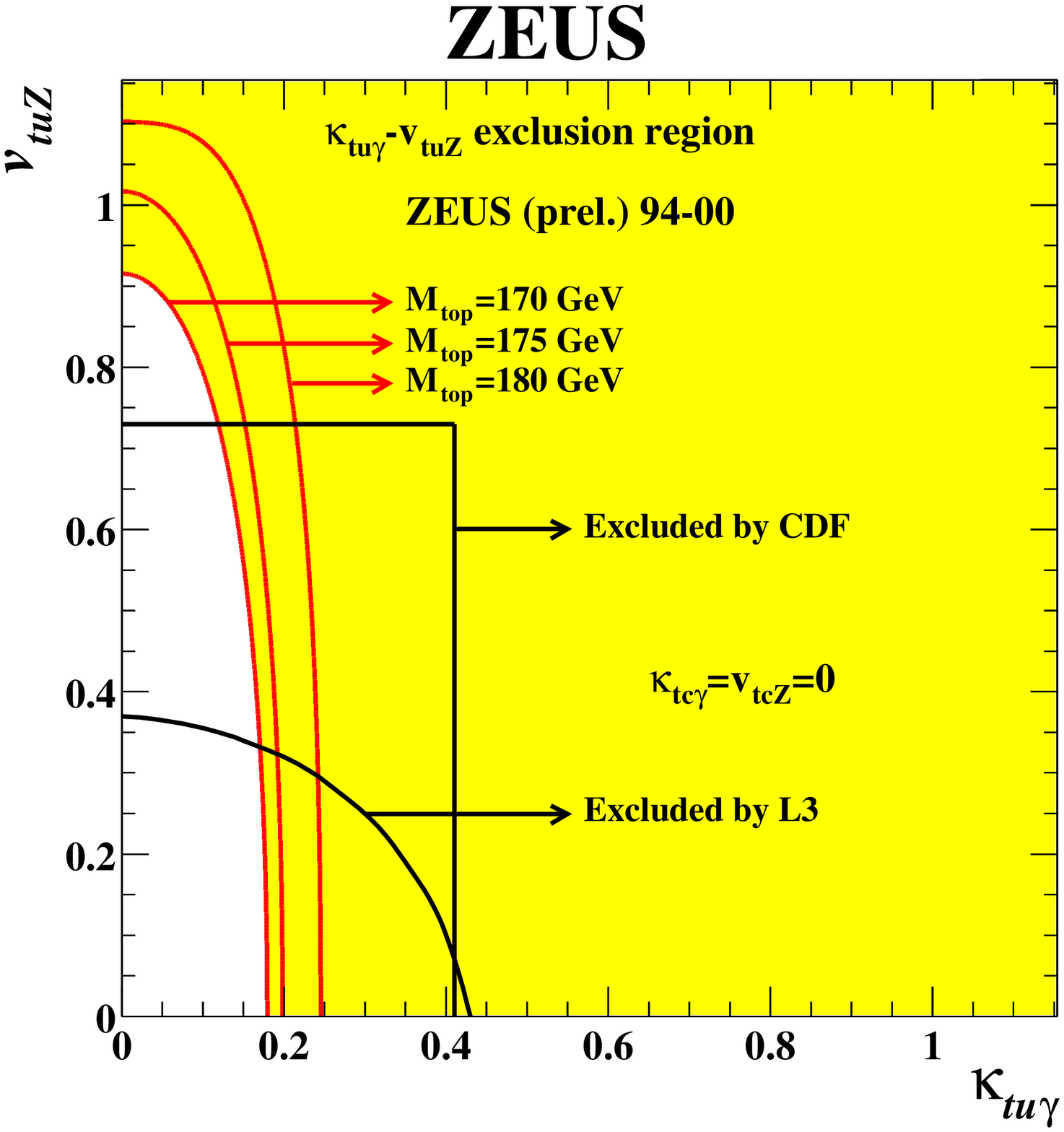,
width={7.4cm}, angle=0, clip=}
\caption[*]{Exclusion regions at 95\% C.L. in the $\kg$-$\vz$-plane from L3, CDF and ZEUS.}
\label{fig-fcnclimits}
\end{minipage}\hfill
\end{center}
\end{figure}

The results from the searches in both the leptonic and hadronic $W$ decay
channels were combined to constrain the production of single-top quarks 
through FCNC. 
The 95$\%$ confidence level (CL) limit for
$\kg$, evaluated assuming $\vz=0$ and $M_\mathrm{top}=175$~GeV, is 
$\kg < 0.174$. 
Next to leading order (NLO) QCD corrections to the single-top 
production cross section~\cite{belyaev} have been taken into account.
In addition, the effect of a non-vanishing $\vz$ coupling was considered. 
In this case a leading 
order cross section was used.
The LEP \cite{l3a} and Tevatron \cite{Tevatron} experiments 
have performed similar searches for single-top
production and rare top decays, respectively. They are sensitive to anomalous
top-quark couplings to $u$ and $c$ quarks through both the photon and the $Z^0$ boson.
The results of these searches are compared to the ZEUS limits in 
Fig.~\ref{fig-fcnclimits}. The ZEUS limits are shown for three different values of 
$M_\mathrm{top}$, since the uncertainty on the top mass is the dominating systematic
uncertainty.
It is evident that HERA is competitive in searches for FCNC couplings
through photon interactions.

\end{document}